\documentclass[journal]{IEEEtran}

\usepackage{tikz}
\usepackage{amsmath}
\usepackage{tabularx}

\usepackage{float}
\newfloat{algorithm}{t}{lop}
\usepackage{array}
\usepackage{booktabs}
\usepackage{url}
\usepackage{color}
\usepackage{multirow}
\usepackage{siunitx}
\usepackage[a-1b]{pdfx} 

\usepackage{caption}
\usepackage{subcaption}

\usepackage{textcomp}

\newcount\Comments  
\usepackage{color}
\definecolor{darkgreen}{rgb}{0,0.5,0}
\definecolor{darkblue}{rgb}{0,0,0.5}
\definecolor{purple}{rgb}{1,0,1}

\newcommand{\kibitz}[2]{\ifnum\Comments=1\textcolor{#1}{#2}\fi}

\newcommand{\George}[1]{\kibitz{darkblue}      {[GG: #1]}}

\begin{document}

\date{}

\title{\Large \bf Compromised ACC vehicles can degrade current mixed-autonomy traffic \\ 
performance while remaining stealthy against detection.}



\author{George Gunter, Huichen Li, Avesta Hojjati, Matthew Nice, Matthew Bunting,\\ Carl A.~Gunter, Bo Li, Jonathan Sprinkle, Daniel Work
\thanks{G. Gunter, M. Nice, and D. Work are with the Department of Civil and Environmental Engineering and the Institute for Software Integrated Systems, Vanderbilt University, TN. email: (\{george.l.gunter, matt.nice, dan.work\}@vanderbilt.edu).}
\thanks{J. Sprinkle is with the Department of Electrical and Computer Engineering and the Institute for Software Integrated Systems, Vanderbilt University, TN. email: (jonathan.sprinkle@vanderbilt.edu)}
\thanks{H. Li, A. Hojjati, C. A. Gunter, and B. Li are with the Department of Computer Science, University of Illinois at Urbana-Champaign, IL, email: (\{huichen3,hojjati2,cgunter,lbo\}@illinois.edu).}
\thanks{M. Bunting is with the Department
of Computer and Electrical Engineering, University of Arizona, AZ, email: (mosfet@arizona.edu).}
\thanks{*Corresponding author is G. Gunter: george.l.gunter@vanderbilt.edu}}%



\maketitle

\begin{abstract}
We demonstrate that a supply-chain level compromise of the \textit{adaptive cruise control} (ACC) capability on equipped vehicles can be used to significantly degrade system level performance of current day \textit{mixed-autonomy} freeway networks. Via a simple threat model which causes \textit{random deceleration attacks} (RDAs), compromised vehicles create congestion waves in the traffic that decrease average speed and network throughput. We use a detailed and realistic traffic simulation environment to quantify the impacts of the attack on a model of a real high-volume freeway in the United States. We find that the effect of the attack depends both on the level of underlying traffic congestion, and what percentage of ACC vehicles can be compromised. In moderate congestion regimes the attack can degrade mean commuter speed by over 7\%. In high density regimes overall network throughput can be reduced by up to 3\%. And, in moderate to high congestion regimes, it can cost commuters on the network over 300 $\frac{USD}{km hr}$. All of these results motivate that the proposed attack is able to significantly degrade performance of the traffic network.

We also develop an anomaly detection technique that uses GPS traces on vehicles to identify malicious/compromised vehicles. We employ this technique on data from the simulation experiments and find that it is unable to identify compromised ACCs compared to benign/normal drivers. That is, these attacks are stealthy to detection. Stronger attacks can be accurately labeled as malicious, motivating that there is a limit to how impactful attacks can be before they are no longer stealthy. 


Finally, we experimentally execute the attack on a real and commercially available ACC vehicle, demonstrating the possible real world feasibility of an RDA. In particular, we test two well-known CAN bus detection techniques on the ensuing data and find that they fail to identify the malicious messages, suggesting that the attacks may be stealthy to detection at the vehicular level as well. These results suggest that current-day mixed-autonomy traffic may be vulnerable to cyber-attacks that can degrade system-level performance, all while remaining stealthy against detection.
\end{abstract}

\section{Introduction}\label{sec:Intro}

\begin{figure*}
    \centering
    \includegraphics[width=0.95\textwidth]{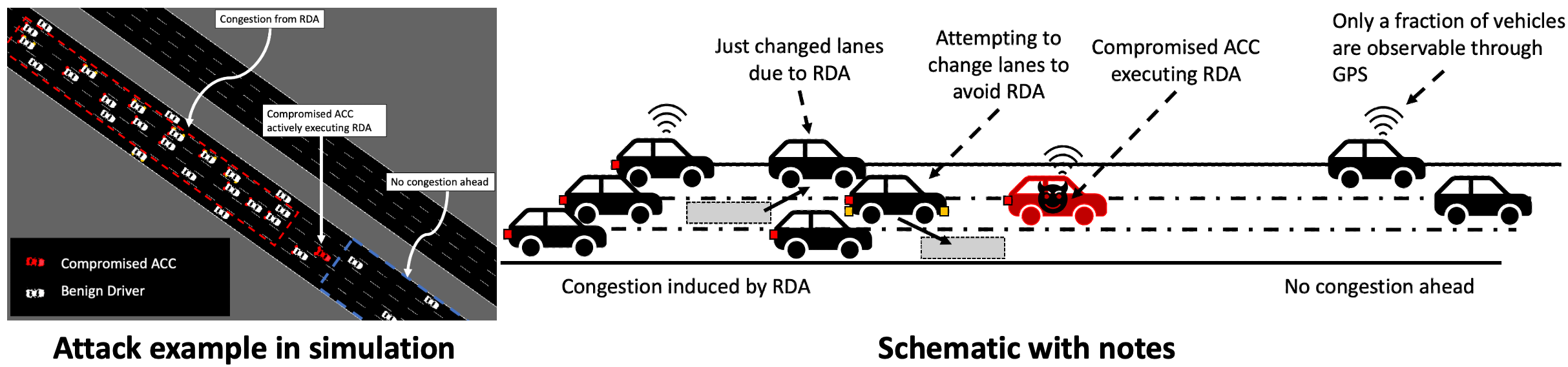}
    \caption{Left: A snapshot of the proposed attack active in our simulation environment causing artificial congestion across multiple lanes. Right: A labeled schematic explaining how the proposed attack can lead to increased traffic congestion, while plausibly being difficult to detect due to sparse GPS measurements.}
    \label{fig:intro_figure}
\end{figure*}

In recent decades, automobile technology has moved significantly in the direction of automating control away from human drivers and moving to \textit{Automated Vehicles} (AVs). While some notable technologies, such as vehicle-to-vehicle and vehicle-to-infrastructure communication, have lagged in commercial feasibility and deployment, \textit{Advanced Driver Assistance Systems} (ADASs) are experiencing wide-spread adoption in commercially available vehicles. \textit{Adaptive Cruise Control} (ACC) is one such technology. ACC controls the speed of a vehicle while actively following vehicles ahead in traffic. It is currently commercially available on many (perhaps the majority) of current best-selling vehicles internationally~\cite{acc-availability}. The consequence of the spread of these technologies is that current-day traffic is of a \textit{mixed-autonomy} nature, meaning that some vehicles are largely controlled by human drivers while other vehicles are significantly controlled by automated systems. Moreover, within those automated systems different vehicles may have different designs. It is an active area of research how this will shape traffic patterns in the future. In Figure~\ref{fig:intro_figure} a preliminary schematic is shown that highlights how compromised AVs within a mixed-traffic environment could negatively impact traffic congestion patterns.

Despite their widespread adoption, it is of significant concern that ADAS features are expanding attack surfaces for AVs~\cite{petit2015potential}. While an increase in the possibility of compromises poses a threat to individual vehicles (i.e. causing a crash), a less studied problem is how it may also introduce a new attack vector by which vehicular traffic systems at large can be attacked. Additionally, transportation networks are crucial to economic operation and development, with congestion costing on the order of billions of dollars annually in the United States~\cite{congestion_costs}. Despite this, relatively little work has been done investigating the vulnerability of current  transportation systems beyond the individual vehicle level.

Moreover, despite the importance of transportation systems, the ability to observe their behavior is currently limited. The most common observation techniques for vehicular traffic are fixed-point sensors, such as inductive loop detectors and radar measurements, which give aggregate metrics of traffic flow performance at points, but no information on the behavior of individual vehicles. GPS measurements from mobile devices or on-board units allow measurements on the driving behavior of individual vehicles, but may be sparsely adopted in the traffic flow and thus provide only partial information about driving behavior. 

\noindent\textbf{Scope.} This paper focuses on threats to current mixed-autonomy vehicular traffic systems in the presence of compromised ACC systems in vehicles. In general: if an adversary were to gain some control of a portion of vehicles running ACC, perhaps through a supply-chain compromise, could attacks that are local to each vehicle combine to form a global attack that is \textit{effective} at degrading overall traffic flow? Additionally, could such attacks also be \textit{stealthy} against detection? This is a common question considered for \textit{Distributed Denial of Service} (DDoS), where adversaries seek a sweet spot that causes maximum damage but avoids exposing attacking machines~\cite{bhuyan2015empirical}. Additionally, other works have explored ways in which large-scale execution of many low-intensity attacks can combine to degrade the performance of a whole system like the power grid~\cite{power_grid_compromise}. The current paper carries the spirit of such efforts into the area of mixed-autonomy vehicular transportation systems. In summary, our concern focuses on the capacity for a cyber-attack to compromise transportation at the \textit{system} level.

\noindent\textbf{Attack model. } We develop and explore an attack which we call a \textit{Random Deceleration Attack} (RDA) that can be executed on a compromised ACC. The attack does not require expert level design (i.e. the attacker does not require knowledge of how the ACC is otherwise controlled). We find this attack is able to affect the system-level performance of a complex and realistic multi-lane traffic environment containing both ACC drivers and human drivers (i.e. mixed autonomy). We focus on this problem because it represents the current operational conditions for vehicular transportation networks. This contrasts with coordinated platoons, which have received much more research attention but do not exist currently in practice. In other words, under current operational conditions, are our transportation systems vulnerable to cyber-attacks?

\noindent\textbf{Defense model.} To determine whether or not the proposed attack is stealthy to detection we formulate an anomaly detection technique which utilizes GPS measurements (from mobile devices perhaps) and attempts to identify compromised vehicles within a broader traffic flow. The anomaly detection is a single-class detection technique based on autoencoders, meaning that it is trained on regular driving patterns and assumes no attack model. Subsequently, attacks that are not classified as malicious are considered to be stealthy to detection.

\noindent\textbf{Contributions.} The contributions of this work are fourfold.
\begin{itemize}
    \item We show that an attack that performs brief overrides to ACC control of a vehicle's speed via a supply-chain compromise to the ACC software is able to significantly degrade the performance of current-day mixed-autonomy traffic at the system level using a simple technique.
    \item Compromised vehicles are not caught by the detection techniques, meaning they are stealthy to detection. Particularly strong attacks may be correctly classified.
    \item We develop a software package built on popular open-source traffic modeling software, with detailed traffic simulation scenarios built into the repository. We pair this with formalizing the detection problem so that other researchers can expand on the work presented in this paper.
    \item On a real commercially-available ACC we experimentally validate that the attack can be executed. Additionally, we test two popular CAN bus detection techniques on data recorded from the vehicle and find that they do not identify the attack as malicious (i.e. it is stealthy).
\end{itemize}

These results and contributions suggest that our vehicular transportation networks have become vulnerable to cyber-attacks that can significantly degrade performance while remaining stealthy to detection.

\noindent\textbf{Organization.} The paper is organized as follows. First, a literature review is done in Section~\ref{sec:related_works}. Next, background and preliminary material on traffic simulation and anomaly detection is covered in Section~\ref{sec:background}. The attack model is described mathematically in Section~\ref{sec:attack_model}. Section~\ref{sec:defense_model} details the defense model used. Subsequently, the impact of different attacks on a detailed/realistic traffic model are quantified in Section~\ref{sec:attack_results}. The stealth of the proposed attack is quantified in Section~\ref{sec:attack_detection}. Experimental work on a live ACC is presented in Section~\ref{sec:vehicle_experiments}. The work is discussed in Section~\ref{sec:discussion}, and concluded in Section~\ref{sec:conclusions}. 

\section{Related works}\label{sec:related_works}
In this section we review related works considering cyber-attacks on AVs and transportation systems. Attacks both on infrastructure, and on vehicular traffic are considered. Anomaly detection in transportation is discussed.

\noindent\textbf{Growing attack surfaces on AVs.} In 2017 a group of researchers successfully implemented a remote attack on the Tesla Model S in both parking and driving modes~\cite{nie2018over}. What made this attack different than the highly-publicized remote exploitation that took place by Miller and Valasek~\cite{miller2015remote} was its ability to remotely compromise many in-vehicle systems such as gateway and autopilot ECUs (which included ACC capabilities). In particular, it enabled malicious injection of messages into the CAN Bus. Tesla was able to fix these vulnerabilities in a relatively short period of time, but in the meantime, there was a fleet-level vulnerability.  

Compromising individual vehicles via physical access or remote capabilities is not new. Checkoway et al.~\cite{checkoway2011comprehensive} discovered that remote exploitation is feasible via a broad range of attack vectors. Most of these vulnerabilities are related to the CAN bus protocol, which lacks security protections such as encryption and authentication~\cite{bozdal2020evaluation}. Additionally, expansions to vehicular communication present another avenue for compromise~\cite{dibaei2019overview}. Arguably, the next generation of AVs is likely to be at even higher risk to these types of attacks, possibly at wide-scale compromise, all while more control is shifted away from drivers to automated systems on the vehicle.

\noindent\textbf{Attacks on transportation infrastructure. }
Original work on security issues in transportation focused on attacking infrastructure systems. In~\cite{ghena2014green} it was demonstrated that traffic lights could be infiltrated and manipulated to favor certain parties. \cite{laszka2016vulnerability} showed that compromises to traffic lights could also be used to generate artificial congestion throughout a network. Similarly, \cite{reilly2015cybersecurity} showed that compromised ramp-meters on freeways could be used to create increased congestion. At the intersection between attacking vehicles and infrastructure \cite{chen2018exposing} demonstrated how in an environment with connection between vehicles and infrastructure, a set of compromised vehicles could be used to attack a connected traffic light in order to create artificial congestion.

\noindent\textbf{Attacks on individual vehicles.} Of recent interest are attacks on individual AVs. In \cite{petit2015potential} a number of attacks were demonstrated that performed remote compromising of camera and lidar sensors on AVs, and in \cite{cao2019adversarial} these were extended using adversarial analysis techniques.~\cite{sitawarin2018darts} demonstrated a capacity for alterations to road-side signs to adversely affect the perception of AVs in an adversarial manner. The work in~\cite{RL_trojan_traffic_attack} found that the use of Trojan attacks on reinforcement-learned AV controllers could be used to cause the vehicle to execute undesirable manoeuvres that endanger the AV. 

\noindent\textbf{Attacking traffic systems.} While many works have looked at the consequences of a compromise of a AV as it pertains to that of the individual vehicle, other works have examined how \textit{platoons} of vehicles could be affected. A platoon of vehicles typically consists of a set of AVs all running a shared control scheme immediately following one another. In~\cite{String_Stability_Vehicular_Platooning,dunn2015attacker,Attack_Change_Control} attacks on CACCs were explored, where it was shown that if a malicious actor with knowledge of the underlying platoon control scheme could gain access to changing the control parameters on a few vehicles in a CACC platoon that the whole platoon could be turned from \textit{string-stable} to \textit{string-unstable}, without needing to compromise other vehicles. This meant that a common design goal (string-stability) could be effectively subverted, and this attack resulted in increases in platoon fuel consumption as well as decreases in vehicular throughput. In other words, an attacker needs access only to a small percentage of agents in the traffic flow in order to attack the performance of the larger system.

Of particular note for the work presented in this paper, the work in~\cite{Attack_Change_Control,dunn2015attacker} also showed that the same attack framework could be adopted for attacking ACC platoons (rather than CACC). Across a stream of vehicles all running the same ACC control scheme, which was otherwise string stable, it was shown that a small penetration rate of compromised ACCs could also be used to turn the platoon from string-stable to string-unstable. Additionally, the traffic achieved worse performance in throughput and speed than if the traffic at large had otherwise been human controlled, suggesting that compromised autonomy may be worse than no autonomy. This lays the groundwork for the work we propose here, in that it motivates that a small percentage compromise of ACCs can be leveraged to impact global traffic performance. 
 
We identify four key ways in which the attacks in~\cite{Attack_Change_Control,dunn2015attacker} can be extended to closely represent current day traffic operating scenarios. First, our attack does not focus in particular on disrupting string stability, a property that has recently been shown not to hold for any commercially available ACC systems~\cite{gunter2019are,gunter2019model,milanes2014modeling}. Rather, our attack just focuses on degrading more general metrics of performance in a traffic flow, such as speed and vehicular throughput. Second, our attack is executed in a \textit{mixed-autonomy} environment in which both human and ACC drivers are present. Currently platooning of CACCs/ACCs is not commercially deployed, while individual vehicles running ACC within the borader traffic flow are. Third, we assess the attack on a multi-lane traffic environment with complicated congestion dynamics and a real world network geometry, not in a single-lane scenario. Finally, and most importantly, our attack requires no knowledge on the part of the attack designer of either the ACC control parameters, nor the dynamical model parameters for other vehicles in the traffic flow for it to be successful in degrading system performance. In other words, the attack does not require an expert attack designer, but is general in nature and easy to implement given adequate access to the vehicle. Across all of these efforts we attempt to move the analysis towards assessing the vulnerability of current day traffic. 

\noindent\textbf{Anomaly detection for transportation systems}
A complete review of general anomaly detection is not in the scope of this work. As a result we will cover works related to anomaly detection in transportation systems. In \cite{urban_traffic_anomalies} two types of detection problems are proposed: in a \emph{flow} detection problem a defender looks for anomalies in aggregate metrics of a transportation system, while in \emph{trajectory} detection we look for anomalous behavior from individual agents. In \cite{flow_detection,flow_detection_dirichlet,flow_detection_guassian} a variety of statistical approaches are developed for finding anomalies in time series measurements of freeway traffic flow, primarily motivated by the need to identify when an accident has occurred. As examples of trajectory detection \cite{taxi_outliers_1} and \cite{taxi_outliers_2} developed techniques for using GPS traces of taxi drivers to determine when anomalous routes were taken. 

In the vein of the works which explore compromised platoons of CACCs and ACCs, recent work has also examined ways in which to detect such intrusions.~\cite{dadras2018reachable_set_adversarial} developed a framework for detecting anomalous behavior in a platoon through the use of reachable set analysis. By developing an invariant set representation of the platoons behavior under standard operation, malicious actors can efficiently be detected if they stray from that space. Additionally in~\cite{dadras2018identification} system identification is performed on ACC/CACC system to find a representative model, from which both spectral and time-domain techniques can be used for anomaly detection. Both methods were computationally efficient and could identify individual malicious actors, rather than simply labeling the entire platoon as anomalous.

While those works build on using expected models of platoon behavior a different approach is to use \textit{deep-learning} techniques to perform detection. In~\cite{khanapuri2019learning} and~\cite{khanapuri2021learning} this is done via supervised training of both feedworward neural networks and convolutional neural networks on data from both attacked and unattacked platoons. These techniques use full observation of the platoon including each vehicle's speeds and inter-vehicle spacings to perform anomaly detection.

\George{This and following paragraph should be combined. }We adopt a detection strategy using a deep-learning approach, rather than the model-driven approach, in this paper. However, two main differences exist between this work and the work in~\cite{khanapuri2019learning} and~\cite{khanapuri2021learning}. First, we employ a semi-supervised learning approach in which we assume the defender has access to training data that is benign in nature (i.e. non-attacked), but has no knowledge of what an attack may look like. Secondarily, we assume that the defender has access only to GPS measurements from individual vehicles. This mean the defense only has access to vehicular speeds, and not inter-vehicle spacings, as is assumed in~\cite{khanapuri2019learning,khanapuri2021learning}. This is because GPS measurements cannot be used to directly find inter-vehicle spacing values unless it is already known that one vehicle is following another, which in practice is not the case. 

We choose this detection scenario as it represents the fact that vehicular traffic network operators currently have access either to only aggregated measurements (average speeds and vehicular counts), which cannot be used to identify individual malicious agents, or they have access to GPS traces from vehicles, possibly coming from mobile devices~\cite{mobile_century}. While in a few experimental scenarios over-head cameras on poles or drones have been used to collect complete traffic information~\cite{HighD_Dataset,NGSSIM} (both speeds and inter-vehicle spacing), current traffic monitoring systems do not have this level of system observation. Again, in this work we attempt to create a scenario for analysis that represents current operational conditions.

\section{Preliminaries and Background}\label{sec:background}
In this section background and preliminary topics are covered. First, we give an overview of how traffic is modeled when individual vehicles are considered. Next, we describe anomaly detection, and in particular how it can be phrased in the context of identifying malicious agents within a traffic flow.

\subsection{Modeling mixed-autonomy traffic}\label{sec:traffic_modeling}
To understand the consequences of attacks on vehicles that can disrupt traffic overall, mathematical models of traffic are needed. In this work a \textit{microscopic} traffic simulation approach is used in which individual vehicles are simulated using \textit{ordinary differential equations} (ODEs). Each vehicle is described via a set of ODEs that dictate its motion as a function of the vehicle's state and the state of the driver ahead of it. In the transportation literature, such ODEs are known as \textit{car following models} (CFMs), and frequently take the form:
\begin{equation}\label{eq:generic_CFM}
\begin{array}{rl}
& \dot{s}(t) = \Delta v(t),\\
& \dot{v}(t)= f(s(t), v(t), \Delta v(t)), 
\end{array}
\end{equation}
where $s(t), v(t)$, and $\Delta v(t)$ correspond respectively to the space-gap (distance) to the lead vehicle immediately ahead, the speed of the ego vehicle (that is, the vehicle whose motion is being described), and the difference in speed between the ego vehicle  and the lead vehicle. The CFM $f(\cdot, \cdot, \cdot)$ describes the acceleration $\dot{v}(t)$ of the ego vehicle based on these quantities. In this work different CFMs are used to model human drivers, normal ACCs, and compromised ACCs. By using these CFMs to represent many vehicles along a freeway network we can perform simulations to understand the consequences of attacks on ACC vehicles on traffic systems.

\noindent\textbf{Modeling CAVS in traffic. }A common CFM to describe ACC vehicles, and the one used in this work, is:
\begin{equation}\label{eq:ACC_CFM}
f_{\text{ACC}}(s,v,\Delta v) = k_{1}(s - t_{h}v) + k_{2}(\Delta v),\\
\end{equation}
where $k_{1}$ and $k_{2}$ are control gain parameters that balance terms to achieve the desired time gap $t_{h}$ and match the velocity of the vehicle ahead (i.e. $\Delta v=0$). We use the corresponding model parameters found in~\cite{gunter2019are,gunter2019model,milanes2014modeling} which were calibrated on real ACC vehicles and have been shown to accurately model the driving behavior of commercial ACCs.

\noindent\textbf{Modeling humans in traffic. }To model human drivers we use the popular \textit{Intelligent Driver Model} (IDM)~\cite{IDM_original}. This ODE takes the following form:

\begin{equation}\label{eq:IDM_CFM}
    f_{\text{IDM}}(s,v,\Delta v) =a\left[1-\left(\frac{v}{v_{0}}\right)^{\delta}-\left(\frac{s^{*}\left(v,\Delta v\right)}{s}\right)^{2}\right],
\end{equation}
where $s^{*}\left(v,\Delta v\right)$ is defined as
\begin{equation}\label{eq:IDM_Spacing_Equation}
    s^{*}\left(v,\Delta v\right)=s_{0}+vT+\frac{\max\{0,v\Delta v\}}{2\sqrt{ab}}\;.
\end{equation}
and $[a,b,\delta,v_{0},s_{0}]$ are model parameters, which we take values for from the literature~\cite{IDM_original}.

\begin{figure}
    \centering
    \includegraphics[width=1.0\columnwidth]{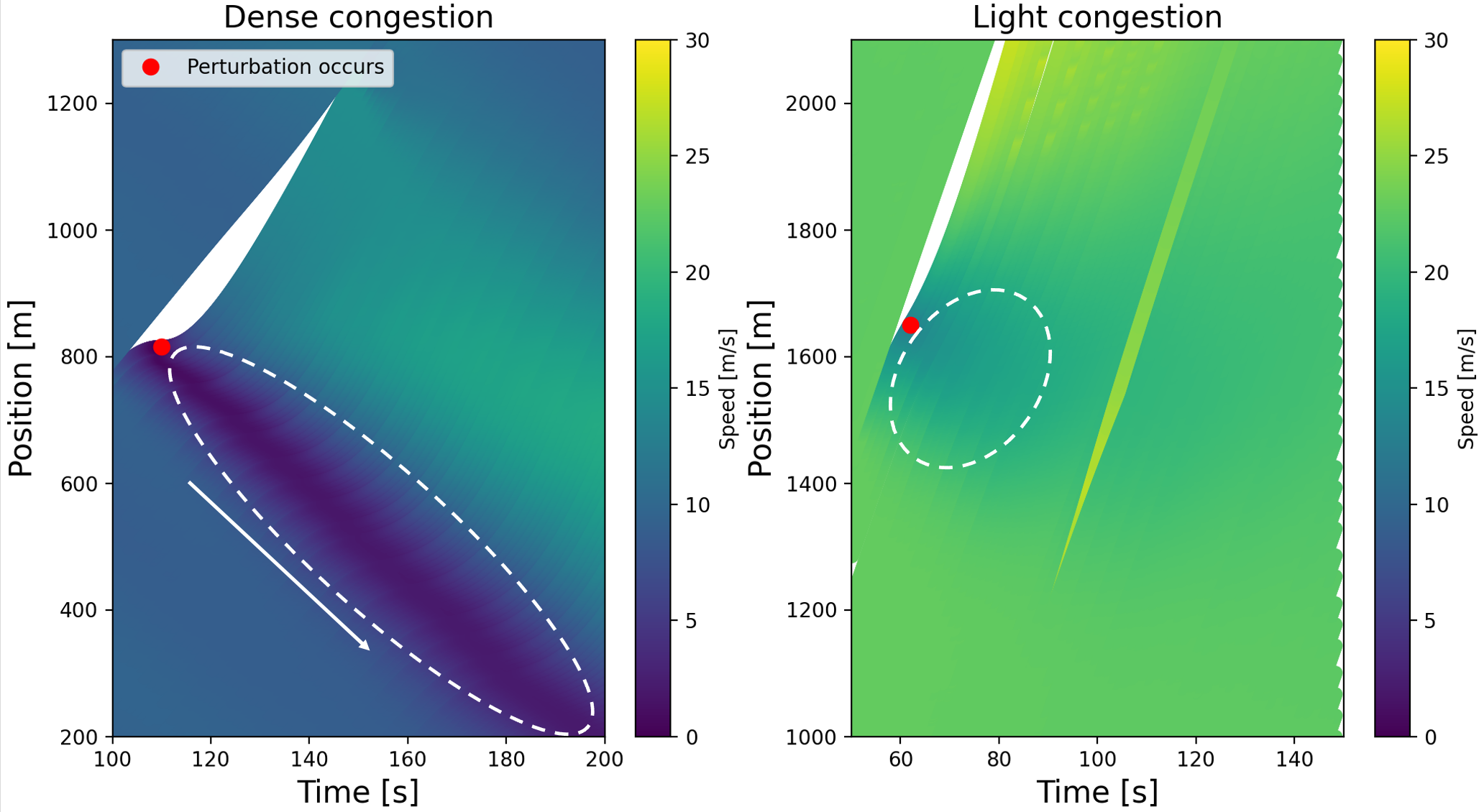}
    \caption{A slow-down event (perturbation) causes a travelling congestion wave to form. In denser traffic the wave grows and moves in the traffic, while in lighter congestion it dies out quickly.} 
    \label{fig:travelling_wave_comparison}
\end{figure}

\noindent\textbf{Traffic congestion patterns. }An important feature of vehicular traffic models is that traffic congestion patterns move both in time and space along a network. Of particular interest for this work are \textit{travelling waves}, which are slow-downs in the traffic flow which translate across vehicles in the flow, and move spatially within the traffic flow. Travelling waves are known to degrade average speed, throughput, and even energy efficiency of traffic flows~\cite{stern2017dissipation}. Travelling waves can emerge in otherwise smooth (i.e. constant speed) traffic, and are created by vehicles slowing down due to an external disturbance, such as a merging events, road curvature, or road grade changes. In the context of this work, our proposed attack also represents a perturbation that could provoke the emergence of traveling waves.


Additionally, travelling waves can have varying speeds, widths, magnitudes, and growth patterns depending on what the underlying density of the traffic flow is, with light traffic tending not to develop waves, while denser traffic flows may. To demonstrate, two illustrative examples are shown in Figure~\ref{fig:travelling_wave_comparison} where two instance of single-lane traffic flow are modeled. On the left densely congested traffic is modeled, and on the right lighter congestion traffic. To view the traffic flow as a whole each vehicle's positions is plotted over time and subsequently colored with respect to its speed. These types of plots are known as 'space-time diagrams'. In each example one vehicle causes a 'perturbation' by reducing its speed by 10 $\frac{m}{s}$ over 10 seconds, which is denoted by the red dot. In the denser regime this creates a travelling wave which grows in size and moves backwards against the traffic flow. In the lighter traffic example the wave quickly disperses, not leading to any significant slow down in the traffic.

\noindent\textbf{Software and network modeling. }An important contribution of this work is that we consider attacks on traffic networks that display complicated and realistic congestion patterns. To do so, we have developed a software package we call \texttt{Anti-Flow}, which is a fork of the  the popular open-source modeling software package Flow~\cite{flow}, which itself is built on SUMO SUMO~\cite{SUMO}. Anti-Flow can be found at \texttt{\url{https://github.com/georgegunter/Anti-Flow}}. Originally, Flow was designed to understand how sparsely adopted AVs could be used to improve overall traffic flow. Anti-Flow examines the adversarial equivalent in terms of how sparsely compromised AVs can hurt traffic.

A core effort of this work is to expand analysis of adversarial attacks on transportation systems to represent current-day operational conditions. We identify the following key features for our framework:

\begin{itemize}
    \item \textbf{Mixed-autonomy.} Both models for human drivers and AVs/ACCs are provided, which is a key component of current traffic networks.
    \item \textbf{Multi-lane environments.} Networks with multi-lane roadways are modeled, on which vehicles can execute both routing-based lane changes and speed-based lane changes (e.g. passing slower cars). Including the capacity for lane-changing creates distinctly different possibilities in congestion patterns compared to single-lane scenarios.
    \item \textbf{Real-world network geometries.} By building Anti-Flow on top of Flow/SUMO users can load in real traffic network geometries via \texttt{Open Street Map}, which can subsequently be used to analyze an attack on real world networks of interest.
\end{itemize}

\subsection{Single-class anomaly detection}\label{sec:anom_detect_background}
\noindent\textbf{Classification. }Anomaly detection presents an approach for defense against cyber-attacks. The goal of an anomaly detection technique is to classify measured samples as either benign/normal or malicious/abnormal. In this work we specifically employ \textit{single-class} anomaly detection techniques. Single-class techniques are classifiers trained only on benign data, but which can be used to identify malicious measurements. Typically, and in this work, this is done by the anomaly detector assigning a scalar value to a new sample $x$, which denotes \textit{how} anomalous that measurements is. If the score is sufficiently high $x$ is considered malicious, and if not $x$ is benign. If we denote the anomaly detector as a function which assigns anomaly scores as $L(x)$, then the classification problem is written as such:

\begin{equation}\label{eq:sample_loss}
\begin{array}{rl}
    & \text{benign if: } L(x) < L_{max},\\
    & \text{else: Anomalous}\\
\end{array}
\end{equation}
where $L_{max}$ is a hyper-parameter that determines the maximum allowed anomaly-score before $x$ is labeled as malicious.  

\noindent\textbf{Models and training. }Subsequently, the design of an anomaly detection technique comes down to how a defender creates the function $L(x)$. A key assumption in this work is that a defender will not have a prior model for what how an attack is structured, but only has access only to benign data coming from observing non-attacked traffic. 

As such, the choice of $L(x)$ must be such that it can be designed without access to both benign and malicious data samples, and thus single-class, or semi-supervised, methods are needed. In this work we use deep autonencoder neural networks to perform detection. The training routine for building an autonencoder anomaly detector consists of solving the following optimization problem:

\begin{equation}\label{eq:ae_loss_function}
\begin{array}{rl}
& \underset{\theta}{\text{min }} \frac{1}{N}\overset{N}{\underset{i=1}{\sum}}L(\theta,x_{i}),\\
& L(\theta,x_{i}) =  (F_{\theta}(x_{i}) - x_{i})^{2}\\
\end{array}
\end{equation}
where $F_{\theta}(x_{i})$ is a given autoencoder that accepts an input $x_{i}$ and is parameterized by network weights $\theta$. Each training sample $x_{i}$ comes from a set of $N$ training samples. $L(\theta,x_{i})$ is the function which assigns a anomaly score to a given sample, which in this case consists of the \textit{squared-error} (SE) between the original sample ($x_{i}$), and it's value after being passed through the autoencoder. $L(\theta,x_{i})$ is frequently referred to as the \textit{reconstruction error} on a given measurement, which when high denotes a suspected malicious measurements. Solving the training problem consists of minimizing the \textit{mean-squared-reconstruction-error} (MSRE) over all training data samples. Other loss functions could be feasibly employed, but the MSRE is common.

\section{Attack model}\label{sec:attack_model}
Here a description of the attack model is given. First, we describe the attack model at the vehicular level in terms of compromises to ECUs and CAN bus messages. Next, the attack model is described at a traffic system level in terms of a CFM. 

\noindent\textbf{Attack model at the vehicular level. } For this attack we envision a scenario in which the software that runs on an ACC unit is compromised, but only in terms of what acceleration commands are sent to the vehicular communication network. More specifically, the compromised software will randomly, and for some time, overwrite the otherwise specified value and instead command the vehicle to engage in a braking event of some magnitude. We denote this attack a \textit{randomized deceleration attack} (RDA).

\begin{figure}
    \centering
    \includegraphics[width=0.95\columnwidth]{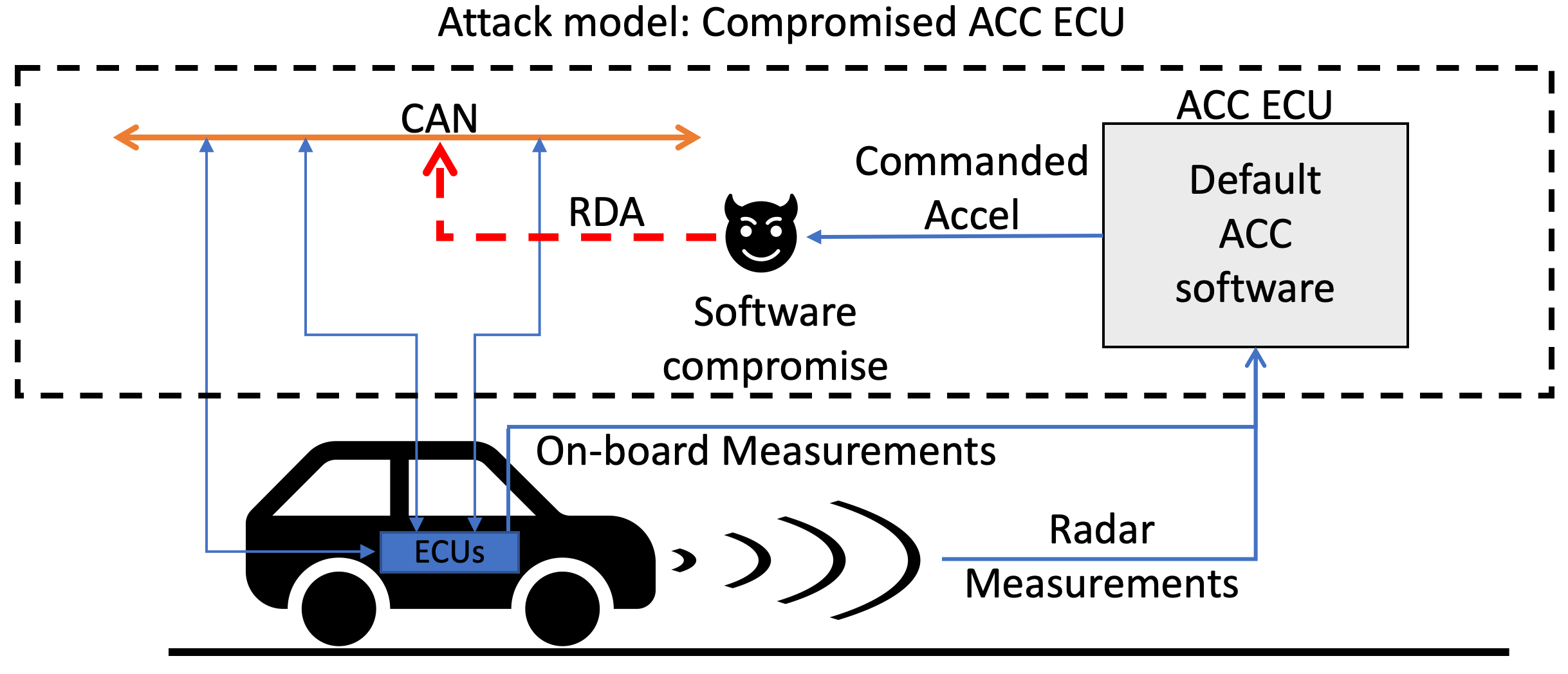}
    \caption{A flowchart demonstrating the specific compromise to the vehicular communication envisioned in this work. A compromise happens on at the level of over writing values sent via the ACC ECU meant to command the acceleration of the vehicle.} 
    \label{fig:vehicle_attack_model}
\end{figure}

In Figure~\ref{fig:vehicle_attack_model} a graphical representation of the attack at the vehicular level is shown. The software that commands the ACC takes in a combination of radar measurements and on-board measurements (such as speed) and uses this information to broadcast a desired acceleration back on to the CAN bus. In our attack model we assume that at this point an RDA can occur when the software compromise overwrites value of the commanded acceleration sent on the CAN bus, but no other information. Importantly, the CAN ID remains the same as it otherwise would, and this attack does not seek to impersonate other electronic control units (ECUs) on the bus, which has been an attack of significant study~\cite{cho2016fingerprinting}. 

\noindent\textbf{Attack model at the traffic level.} Our proposed attack translates to the traffic level by consisting of a subset of ACCs on the road each being forced to engage in RDAs. While any individual RDA may not be particularly impactful, a sufficient compromise of ACCs may lead to system level effects.

To quantify this possibility, we develop a CFM in the style of Equation~\eqref{eq:generic_CFM} that describes how an individual compromised vehicle within the broader traffic flow can execute an RDA within the simulation environment. To execute an RDA two parameters are needed: $a_{\text{attack}}$ and $t_{\text{attack}}$. Here, $a_{\text{attack}}$ is the deceleration that the ACC will execute when the attack is active and $t_{\text{attack}}$ corresponds to the period of time over which the attack is active. We model the driving of a compromised ACC in terms of a CFM as such:

\begin{equation}\label{eq:attack_model}
    f_{comp} = 
    \begin{cases}
    min(a_{\text{attack}},f_{\text{ACC}}) & \text{if  attack and }v\geq 0\\
    0.0 & \text{if attack and }v= 0\\
    f_{\text{ACC}} & \text{if no attack}\\
    \end{cases}
\end{equation}

where $f_{\text{ACC}}(s,v,\Delta v)$ is as described in Equation~\eqref{eq:ACC_CFM}. Compromised ACCs follow standard control from the underlying software (modeled via $f_{\text{ACC}}(s,v,\Delta v)$) whenever the attack is dormant, but execute an RDA when the attack is active. The RDA causes the compromised ACC decelerate constantly at the rate $a_{attack}$ for a time of $t_{\text{attack}}$ seconds. If the vehicle comes to a complete stop then the deceleration is not executed, but rather the vehicle stays stationary. The minimum between $a_{attack}$ and $f_{\text{ACC}}(s,v,\Delta v)$ is chosen when an attack is active so as to avoid crashing the vehicle in the event that the ACC system commands a stronger braking event than the RDA. Collisions should be avoided from the attacker perspective to remain stealthy.

\section{Defense model}\label{sec:defense_model}
In this section we describe how anomaly detection of malicious driving is done. First, a model for how measurements of vehicle's driving behavior is proposed. Next, using the single-class autoencoder detection scheme outlined in Section~\ref{sec:background} we describe how detection of malicious agents within the traffic flow is performed.

\noindent\textbf{Traffic observation. }To detect malicious driving behavior from individual vehicles it must be possible to gather data at the individual vehicle level. We propose for this work that GPS measurements coming from mobile devices which are carried within vehicles can be used for this purpose. This detection model has two benefits. Unlike stationary radar units or inductive loop detecting units which measure aggregated statistics about the traffic flow (flow,density) at individual points along a roadway, GPS measurements provide information at the level of individual vehicles. If GPS measurements from a vehicle are flagged as anomalous, this can be used to identify individual malicious actors, which cannot be done from aggregate statistics. Additionally, GPS measurements from mobile devices could not plausibly be compromised by the attack on the vehicle, unlike perhaps other on-board measurements.

\begin{figure}
    \centering
    \includegraphics[width=0.95\columnwidth]{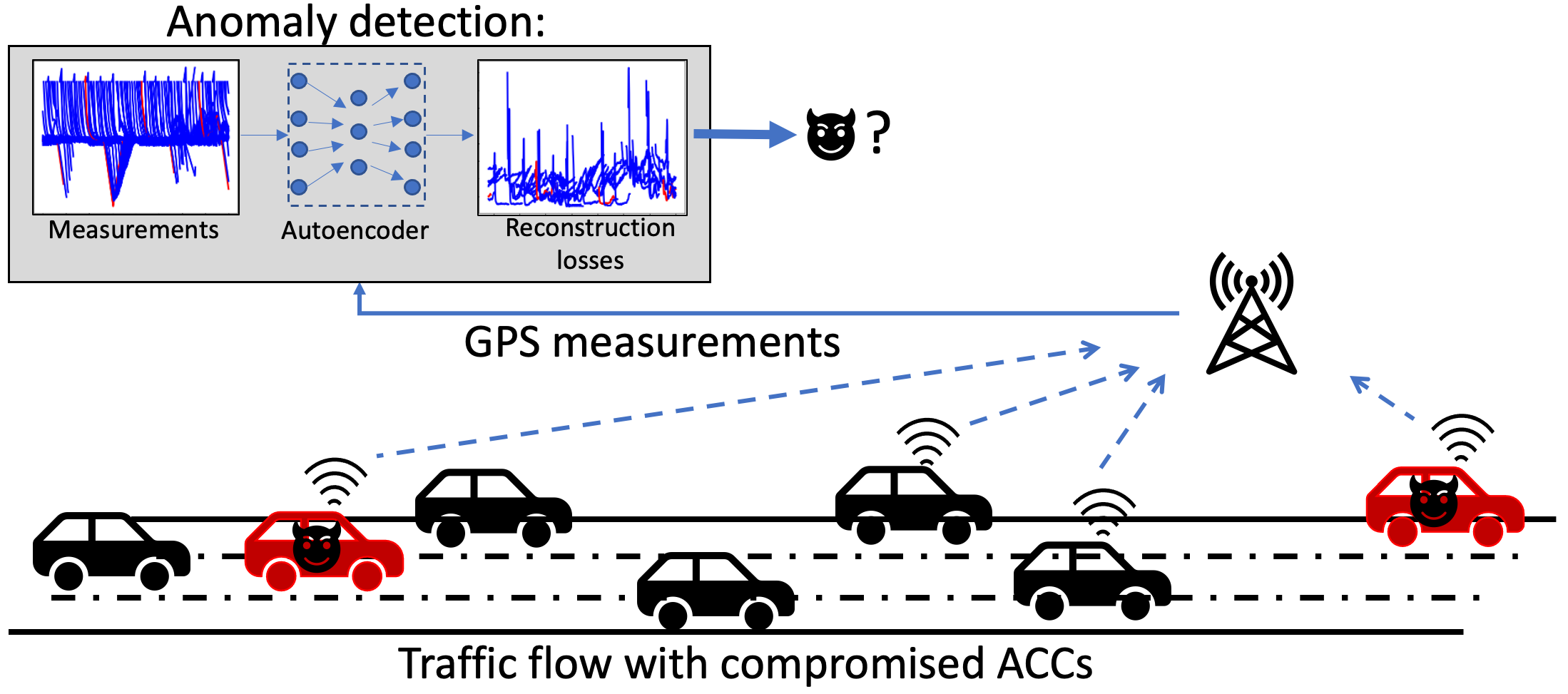}
    \caption{GPS measurements are collected from some vehicles in the traffic flow and subsequently used for anomaly detection.}
    \label{fig:GPS_detect_graphic}
\end{figure}

While in reality only a portion of vehicles within a traffic flow would be equipped with a GPS reporting mobile device, we assume in this work that all compromised ACC vehicles report GPS measurements, and that a portion of all remaining (benign) vehicles also report GPS measurements. This is done to prevent compromised vehicles from passing detection simply because they are unmeasured, although in reality this would be the case. In this work we wish to understand the stealth of the proposed attack even in the context of it being observed.
 
\noindent\textbf{Detection using GPS measurements. } Given the observation model proposed, a defender seeks to identify individual drivers in the traffic flow as malicious or benign. In Section \ref{sec:background} preliminaries on single-class classification of anomalies using autoencoders were covered. Here we adapt those concepts to the problem of identifying malicious agents from their driver behavior.

We assume that the defender is able to gather GPS samples from traffic flows which are un-attacked, and thus all samples can be considered to be benign in nature. In our experimental set up these benign samples come from simulations during which attacks are not present. The architecture we use for our autonencoder consists of a convolutional neural net (CNN), then a long-short term memory (LSTM) recurrent neural net for the encoder portion, and the reverse for performing decoding in a similar style to~\cite{cnn_lstm_timeseries_anomaly} which has been shown to perform well on single-class timeseries anomaly detection. To handle network architecture PyTorch~\cite{pytorch} is used, and each network is trained for 1000 epochs using the loss function defined in Equation~\eqref{eq:ae_loss_function}. A sequence length of 100 is used as input to the autoencoder.

We classify individual vehicles along the network based on anomaly loss scores assigned by the trained autoencoder. In order to develop a loss threshold we use the maximum loss observed on any sample in the training set, as this represents the highest anomaly value known to be benign in nature. Vehicles in the testing regime are assigned an anomaly score by sweeping the autoencoder along the observed speeds and finding the maximum loss. If this value is greater than the threshold found in training, the vehicle is considered anomalous/malicious.

\section{Attacking Traffic}\label{sec:attack_results}
In this section we explore the impact that the proposed attack can have on traffic systems. First, the experimental setup is detailed. Next, we present the results of the numerical experiments run and quantify the impact that attacks have across different traffic conditions.

\noindent\textbf{Experimental overview.} Here we provide a description of the experimental setup. We use a network geometry that is taken from a real United States Interstate Highway and loaded into Anti-Flow using Open Street Map. For simplicity ,only traffic flowing in a single direction is modeled. However, the model provides both on-ramps and off-ramps that enable vehicles to enter and exit the network. The segment of roadway modeled is close to 3 kilometers in length. In reality, this stretch of freeway sees over one hundred thousand vehicles pass along it daily and is therefore of significant economic importance. 

In each simulation traffic is run for 800 seconds with no attacks taking place in order to populate the network and allow it to converge to a regular behavior. This is commonly referred to as the 'warmup' period. Subsequently attacks are allowed, and the simulation is run for 6000 seconds (100 minutes). When calculating impact metrics, only the portion of the simulation involving attacks is used.

\begin{figure}
    \centering
    \includegraphics[width=0.7\columnwidth]{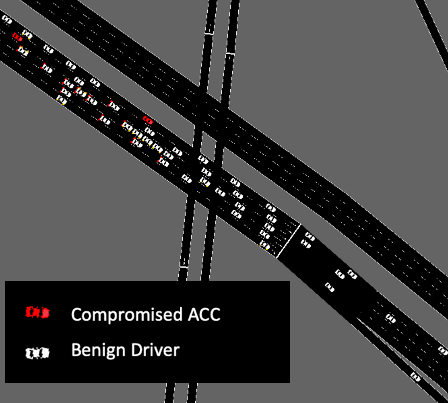}
    \caption{A snapshot of a simulation on the modeled freeway network taken from a real US highway. This freeway has 4-5 lanes and includes both on and off ramps.}
    \label{fig:network graphic}
\end{figure}

We choose two main questions to explore: i) What percentage of ACCs does an adversary need to compromise to significantly degrade traffic performance, and ii) how does the impact of an attack vary across different congestion patterns? To answer these questions we run an ensemble of simulations in which we vary model parameters. We identify 3 key parameters: i) attack strength ii) traffic congestion level, and  iii) ACC compromise rate. Attack strength is specified by selecting values of $t_{\text{attack}}$ and $a_{\text{attack}}$. We select three sets of attacks, which we denote 'weak', 'medium', and 'strong' and have [$t_{\text{attack}},a_{\text{attack}}$] values of [5.0,-0.25],[7.5,-0.5], and [10.0,-1.0]. We also select three congestion levels, which we denote 'light', 'medium', and 'high'. Congestion is set by increasing desired inflow to the network, with higher inflow leading to higher congestion levels. Finally, the ACC compromise rate is varied across values of 5\%,10\%,15\%, and 20\%, while the overall portion of traffic that is made up of ACCs is fixed at 20\%. This means that respectively compromised ACCs will make up 1,2,3, and 4\% of the overall traffic. Baseline simulations in which no attacks are present are run for each congestion level.

\begin{figure}
    \centering
    \includegraphics[width=1.0\columnwidth]{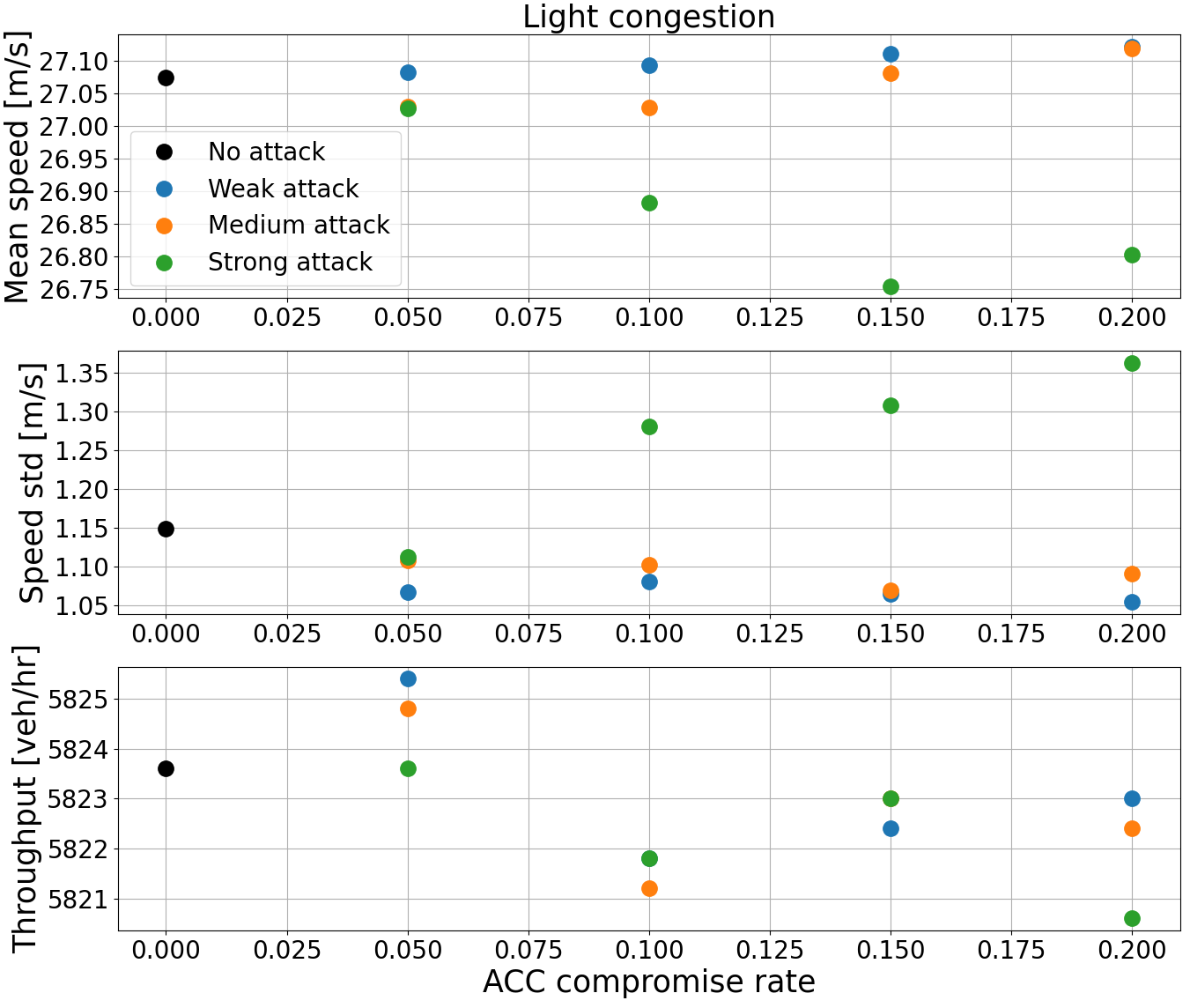}
    \caption{Impact metrics for different attacks/compromise rates in light congestion.}
    \label{fig:light_congestion_impacts}
\end{figure}

\begin{figure}
    \centering
    \includegraphics[width=1.0\columnwidth]{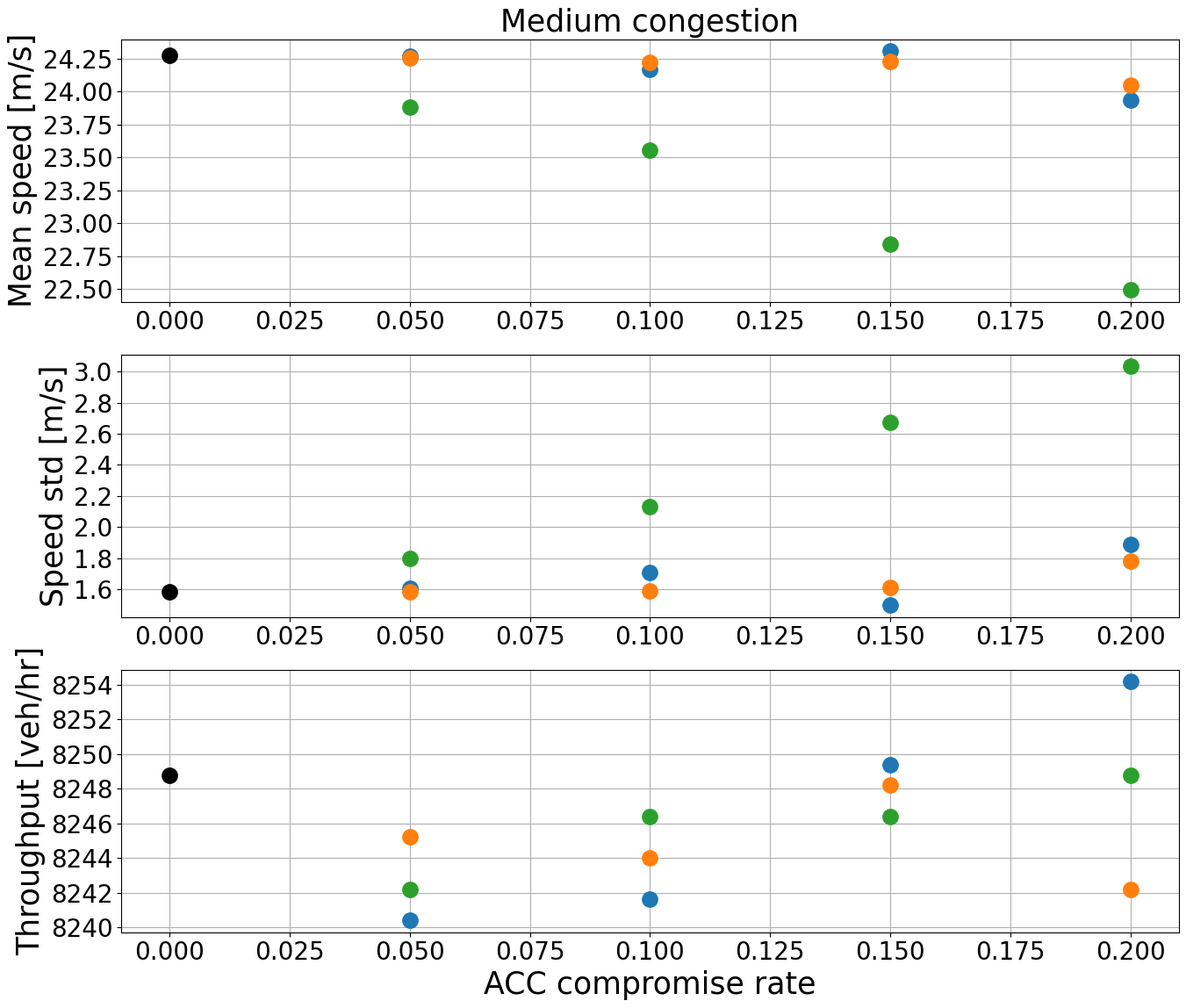}
    \caption{Impact metrics for different attacks/compromise rates in medium congestion.}
    \label{fig:medium_congestion_impacts}
\end{figure}

\begin{figure}
    \centering
    \includegraphics[width=1.0\columnwidth]{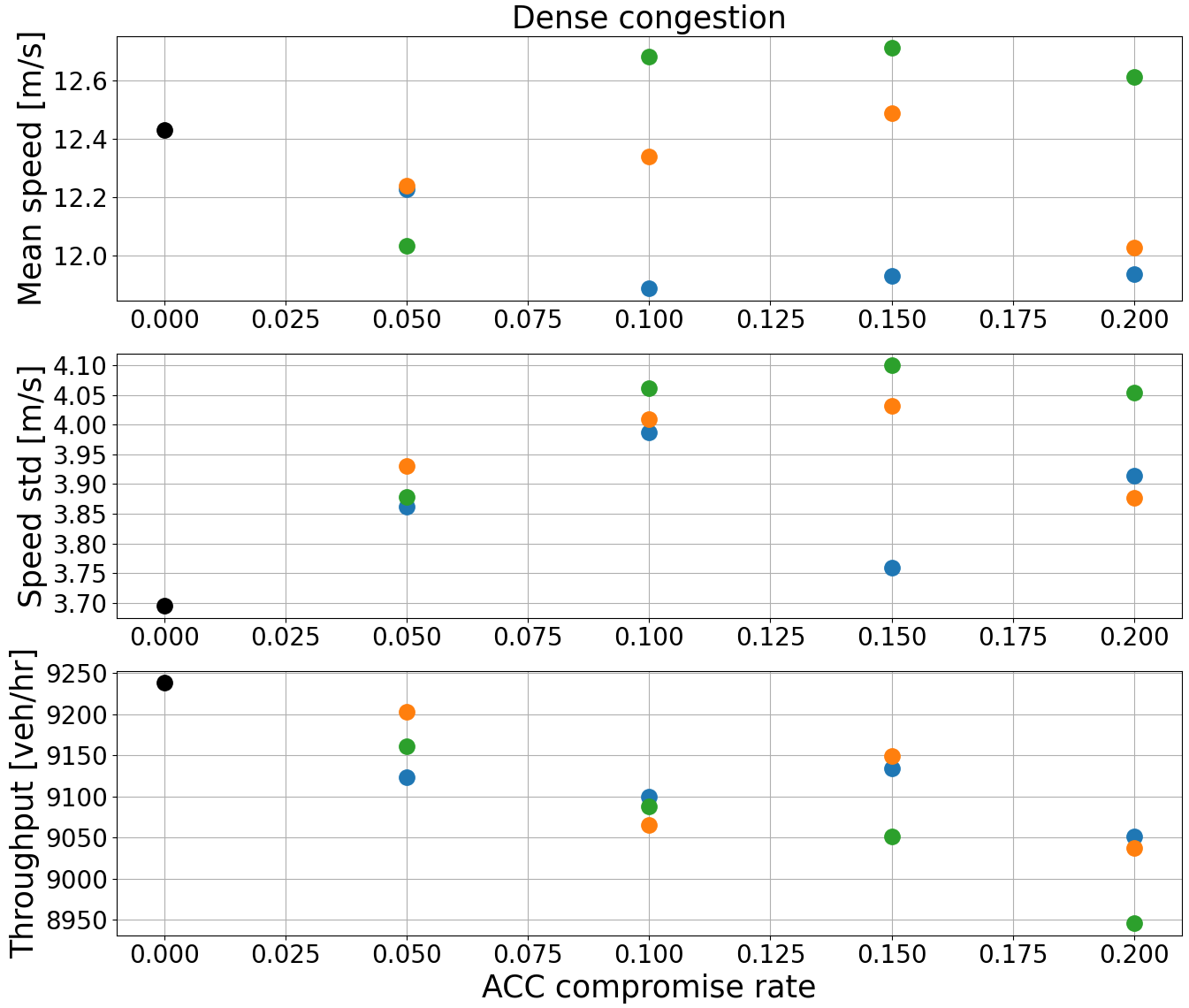}
    \caption{Impact metrics for different attacks/compromise rates in dense congestion.\George{Change to high congestion?}}
    \label{fig:dense_congestion_impacts}
\end{figure}

\noindent\textbf{Impact of attacks.} Simulations measuring impacts of attacks under various conditions are displayed in Figures \ref{fig:light_congestion_impacts}, \ref{fig:medium_congestion_impacts}, and \ref{fig:dense_congestion_impacts}. Three main impact metrics are used to understand the performance of the network, and thus the impact of the attack: mean speed, speed std., and throughput. Mean speed is the average speed that a commuter would travel on the network (e.g. how fast would you expect to drive). Speed std.{} is the average standard deviation in speed a commuter experiences along the network (how much is speed varying during driver). Finally throughput refers to how many vehicles are able to pass through the network per time. Mean speed and speed std.{} are metrics of interest for commuters, while the throughput is a metric of overall efficiency of the system.

In the light and medium congestion regimes the effect that the RDAs have are similar, and most obvious. The weak and medium attacks are mostly ineffectual, while the strong attack is able to decrease the mean speed and increase the speed std.{}, both of which are undesirable. The throughput remains mostly unaffected. Additionally, as the penetration rate of the compromise increases, so too does the impact that the strong attack has on the traffic. From these results we conclude that given a sufficient compromise rate, and attacker would be able to degrade the performance of the network with for commuters in these congestion regimes.

In high congestion traffic, the effects of the attacks are less clear. Unlike the light and medium congestion regimes, all attacks appear be able to effectively degrade the throughput on the network. At a 20\% compromise rate, the the throughput is reduced by 3\%. While this may not seem significant, reductions in throughput on high usage networks can lead to congestion building out into surrounding arterial roadways quickly. In comparison, commuter metrics are not clearly degraded. This may be due to some stochasticity in the simulations, and that further simulations are required to arrive at a definitive understanding. It is also possible that dense traffic already exhibits traveling waves and chaotic dynamical patterns and these cause attacks to be less impactful.

Alternatively, it is also possible that due to the reduction in throughput the underlying congestion rate on the network lessens (i.e. traffic is less dense) leading to an \textit{increase} in mean speed potentially. Further research into utilizing traffic flow theory to understand the impacts of attacks on high congestion traffic regimes is needed.  

Looking across the three congestion regimes, the attacks will have different impacts at different congestion rates. If an attacker wished to focus on degrading commuter metrics this is most possible at medium congestion levels (rather than light or high levels). If the focus is on degrading overall system efficiency (e.g. throughput), then focusing on high congestion traffic conditions would be most impactful.

\noindent\textbf{Estimating the cost of an attack to commuters.} In order to provide further contextualization on the impact of the RDAs on traffic flow, we consider the added cost that commuters along this network would incur from reductions in average speed along the network. In particular, we look at how much the most impactful attack at each congestion level contributes to a reduction in average speed. 

To perform this calculation we use a standard USDOT \textit{value of time} estimate of 14 USD per hour~\cite{USDOT_VOT}, meaning that every hour a commuter spends in traffic is estimated at costing them 14 USD. First the average increase in commute time along the network is found using the mean speed. Next, this is multiplied by the VOT to find an average incurred cost. This is then multiplied by the throughput to find an hourly cost across all commuters, and finally normalized by the length of the network. We refer to this quantity as the \textit{average attack cost} (AAC), which units of $\frac{\text{USD}}{km hr}$.

Across the light, medium, and high congestion regimes we find maximum AACs of 36.0, 307.4, and 433.8 $\frac{\text{USD}}{km hr}$ respectively. Interestingly, while the impact to mean speed in the high congestion regime was not as significant as in the medium congestion regime, the cost to commuters is greater due to the greater number of vehicles passing along the network. Additionally, we note that while the magnitude of cost calculated may not seem large, when considering how this cost would accumulate over a larger section of the network (or other networks), and over a longer period of time, the cost to commuters from this attack could be significant. 

\section{Detecting attacks}\label{sec:attack_detection}
In this section we assess the ability for the anomaly detection technique to find compromised ACCs within a traffic stream. First, the anomaly detection technique proposed in  Section~\ref{sec:defense_model} is used on the results from the round of simulation experiments, and it is found that attacks are all stealthy to detection. We explore the problem further by calculating anomaly loss scores on a parameter sweep of RDAs and find that sufficiently strong attacks can trigger true positive classifications. Finally, we investigate the accuracy of anomaly detection on extreme attacks, and find that the anomaly detection technique is able to find clusters of anomalous vehicles, but is unable to distinguish between perpetrating vehicles and victims. This suggests that accurate detection using GPS measurements alone may be difficult or impossible. 

\begin{figure*}
    \centering
    \includegraphics[width=1.0\textwidth]{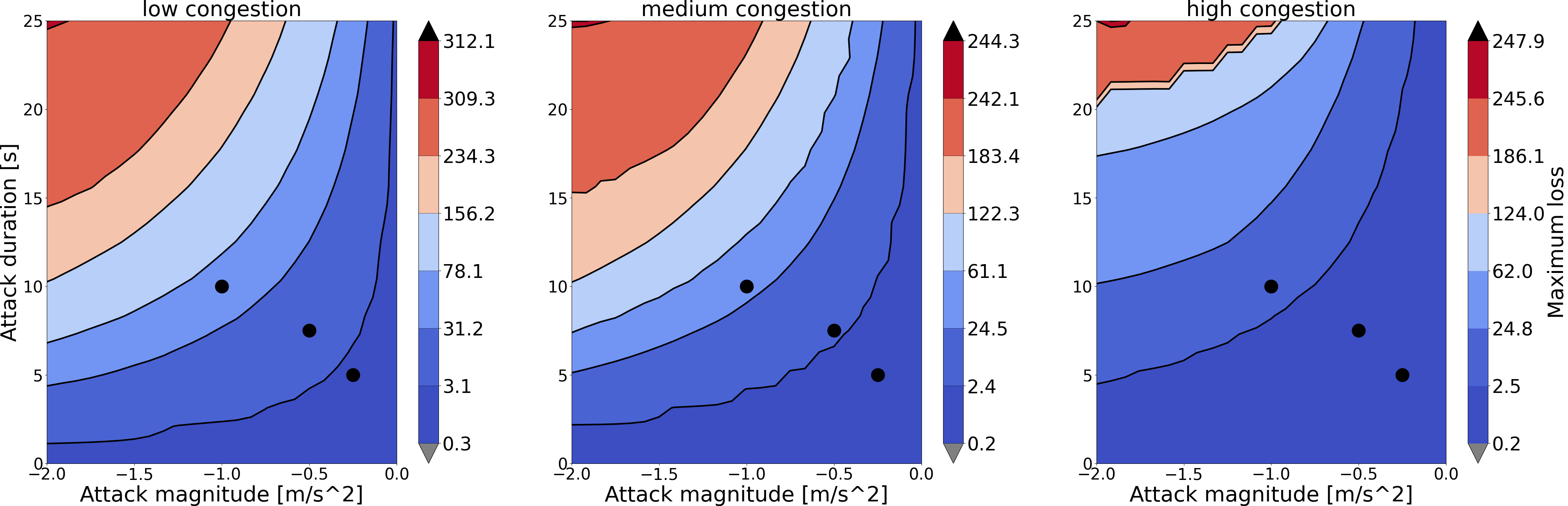}
    \caption{Losses incurred across different RDAs and congestion levels. As RDAs become longer and of greater magnitude, the corresponding anomaly score (maximum loss) increases. Parameter choices for the RDAs in Section~\ref{sec:attack_results} are shown in black.}
    \label{fig:loss_FPR_param_sweep}
\end{figure*}

\noindent\textbf{Could the attacks be detected?} In order to assess the stealth of the RDAs from Section~\ref{sec:attack_results} we now use the autoencoder-based anomaly detection technique proposed in Section~\ref{sec:defense_model} to attempt detection of malicious ACCs that are introducing RDAs. Autoencoders are trained for each congestion scenario on data coming from the benign/unattacked simulations. From each of the baseline simulations 2000 vehicles are selected, and from them 10 random samples are chosen and subsequently used for training. Speed measurements alone are used. The autoencoder structure described in Section~\ref{sec:defense_model} is used in each case.

To perform testing, data is taken from the attacked scenarios. Samples are taken only from the simulation scenario with a 20\% ACC compromise rate since these were the scenarios in which attacks were most impactful. All attack vehicles are tested on, in addition to 1000 randomly selected benign vehicles. This is done to not artificially skew testing rates by selecting significantly more benign vehicles than malicious, as is the case in the real traffic. For each testing vehicles a sweep is done over the time series measurement of their speed and the maximum observed loss across all times is used as the corresponding anomaly score for that vehicle. Subsequently, testing samples are labeled as malicious if the assigned anomaly loss score exceeds the corresponding loss threshold found from the training data.

\begin{table}[]
    \centering
    \begin{tabular}{c|c c c c}
    \toprule
        Congestion/Attack & FPR & TPR & FNR & TPR \\
        \midrule
        Light/Weak & 0.001 & 0.0 & 1.0 & 0.999 \\
        Light/Med & 0.001 & 0.0 & 1.0 & 0.999 \\
        Light/Strong & 0.02 & 0.11 & 0.89 & 0.98 \\
        Med/Weak & 0.004 & 0.011 & 0.989 & 0.996 \\
        Med/Med & 0.0 & 0.013 & 0.987 & 1.0 \\
        Med/Strong & 0.02 & 0.024 & 0.976 & 0.98 \\
        High/Weak & 0.0 & 0.0 & 1.0 & 1.0 \\
        High/Med & 0.004 & 0.0 & 1.0 & 0.996 \\
        High/Strong & 0.001 & 0.0 & 1.0 & 0.999 \\
        \bottomrule
    \end{tabular}
    \caption{Classification rates on the experimental scenarios.}
    \label{tab:attack_classification_results}
\end{table}

In Table~\ref{tab:attack_classification_results} the false-positive rate (FPR), true-positive rate (TPR), false-negative rate (FNR), and true-negative rate (TNR) are shown for all congestion/attack pairs. Across all scenarios the FPR ranges from 0\% to 2\%. While this is a low percentage, 2\% of the overall benign traffic flow is thousands of misclassified vehicles over a day (assuming daily flow on the order of 100,000 vehicles), suggesting that the defender may want to consider an even more conservative loss threshold. Additionally, the TPR is below 3\% across all scenarios, except for the strong attack in light congestion, meaning that in general malicious ACCs are not caught. In the light congestion regime under a strong attack the TPR on the attack is significantly better than in other regimes, suggesting that due to the otherwise calm nature of traffic stronger attacks are more obvious, and as a result an attacker may avoid attacks in those conditions, although most ACCs are still undetected. From the inability for the anomaly detection scheme to properly classify malicious vehicles as such, we conclude that the attacks considered thus far are all stealthy against detection, meaning that there are attacks which are both stealthy and are able to meaningfully degrade the traffic flow.

\noindent\textbf{When are attacks deemed anomalous? }
Given that the attacks examined in Section~\ref{sec:attack_results} are found to be undetectable compared to other benign vehicles in the traffic flow, a reasonable question to ask is whether any RDAs are detectable? To begin answering this, we perform a parameter sweep on the two RDA parameters, but rather than simulate in the context of traffic, we simulate only a compromised ACC performing an RDA by itself. This is done to allow for a computationally efficient way to examine the structure of a wide variety of attacks. We sweep across $a_{\text{attack}}$ in the range of $0.0$ to $-2.0 \frac{m}{s^{2}}$, and $t_{\text{attack}}$ in the range of $0.0$ to $25.0 s$. This is done for each congestion level, where the vehicle is simulated from a starting speed of the mean traffic speed in the unattacked scenario, and then executes the appropriate RDA. From the speed timeseries on the simulations coming from this parameter sweep the corresponding anomaly detector is applied, and the maximum loss calculated, which would be used in the classification problem. The results are shown in Figure~\ref{fig:loss_FPR_param_sweep}.

As the attacks increase in $t_{\text{attack}}$ (brake longer) and decrease in $a_{\text{attack}}$ (brake harder) the maximum loss incurred increases. As a result, given a predetermined classification threshold, the harder braking and longer duration attacks are more likely to be correctly flagged as malicious. This is important, as it suggests there is a cap on how strong an RDA can be made by an attacker before it runs a significant risk of being correctly flagged as anomalous.

\begin{figure}
    \centering
    \includegraphics[width=1.0\columnwidth]{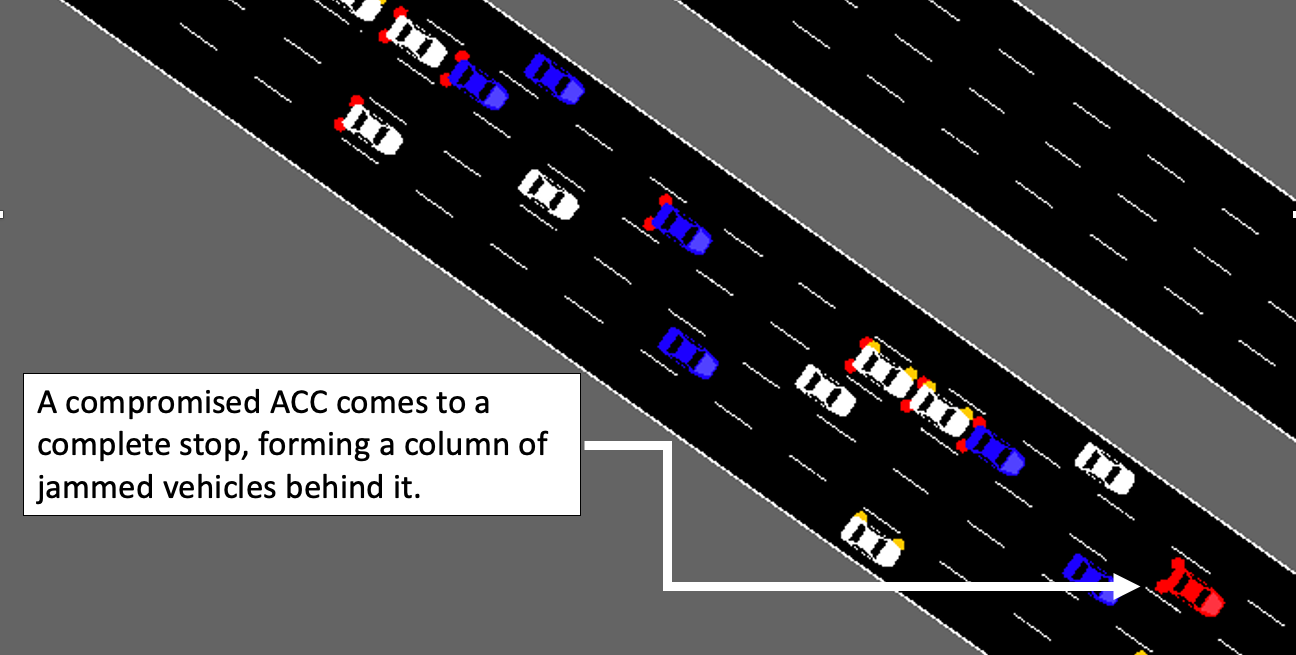}
    \caption{A snapshot of a compromised ACC executing an extremely strong RDA.} 
    \label{fig:ACC_complete_stop}
\end{figure}

\noindent\textbf{Performance of classifiers on strong attacks.}
While stronger attacks may increase the  true positive rate for the anomaly detection, accuracy of the detector is also important. Namely, it may undesirable to misclassify large number benign vehicles in order to correctly flag malicious ones. To investigate this we run another simulation in medium congestion, this time on the full network, and with RDA parameters of $t_{\text{attack}} = 35.0 s$ and  $a_{\text{attack}} = -1.5 \frac{m}{s^2}$. This is an extremely strong attack, and results in a compromised ACC quickly decelerating down to a complete stop, where it remains for some time. A snapshot such an attack is shown in Figure~\ref{fig:ACC_complete_stop}, where a compromised ACC actively executing an RDA is shown creating a column of jammed traffic behind it.

\begin{figure}
    \centering
    \includegraphics[width=1.0\columnwidth]{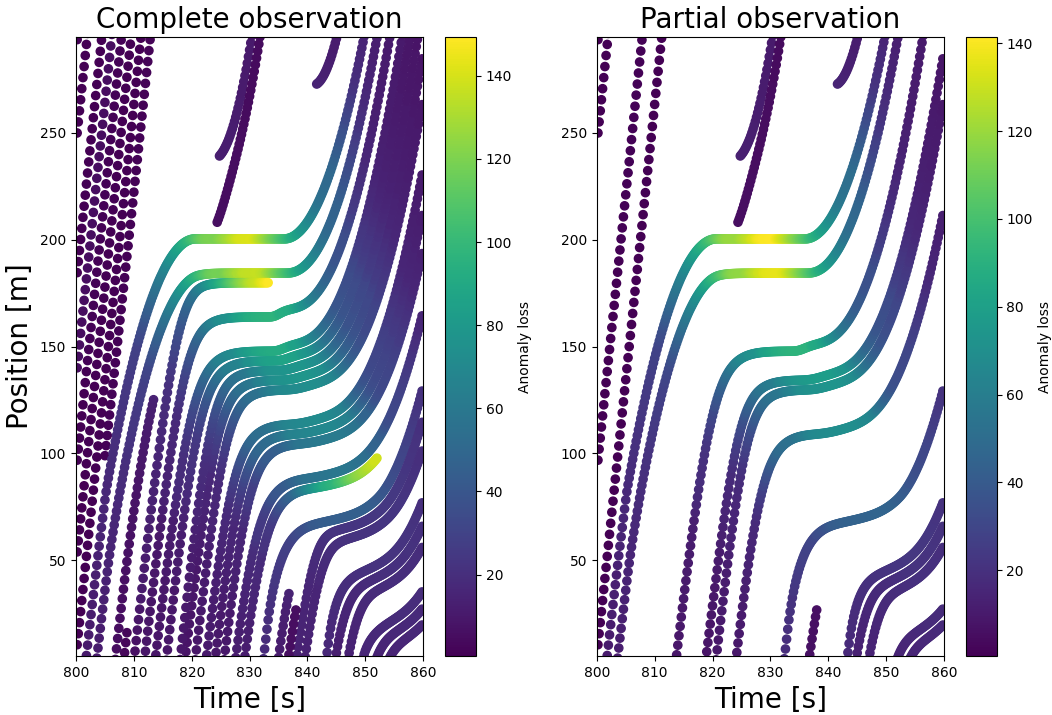}
    \caption{A space-time diagram showing vehicles' trajectories which are caught in the jamming attack. Shading is applied based on anomaly loss values. } 
    \label{fig:anomaly_spread_spacetime}
\end{figure}

In Figure~\ref{fig:anomaly_spread_spacetime} space time diagrams for vehicles involved in the attack are shown. We note that those trajectories which seem to disappear are vehicles changing into an adjacent lane. On the left, trajectories for all vehicles are shown, while on the right only half are shown. In each, coloring is applied in correspondence with anomaly loss scores assigned to the speed profiles of the vehicles over time. 

The left plot shows that with complete observation of the traffic, the attacking vehicle can be identified by looking to the front of the pocket of vehicles all exhibiting high anomaly loss scores. In other words, reasoning about how vehicles are interacting spatially allows for accurate detection. From the scenario with partial observation it is unclear whether an attacking vehicle has been observed, or whether only benign vehicles are being observed as they are part of an anomalous event. As such, reasoning about how vehicles are behaving spatially to one another is impossible due to missing knowledge. While it is possible with the proposed anomaly detection technique to catch sufficiently strong attacks, without the key spatial information that complete observation brings it may be impossible to do this \textit{accurately}.

\section{Analysis on a live vehicle}\label{sec:vehicle_experiments}
A reasonable question to ask at this point is: Could an ACC vehicle with the proposed compromise actually be made to execute an RDA? In this section we experimentally validate the proposed attack by showing that via a compromise similar to the model proposed in Section~\ref{sec:attack_model} (i.e. acceleration commands can be over-written) we are able to make a commercially available ACC vehicle execute an RDA.

We retrofit the vehicle with a CAN bus reading and writing device that allows us to overwrite ACC acceleration commands which were being sent to the CAN bus. We first describe the experimental set up. Next, we show that given this level of access we are able to make the vehicle execute an extreme version of an RDA in which the vehicle slows down for a significant period of time and with significant braking. We collect CAN bus data from a number of experiments involving different RDAs, as well as CAN bus data from normal and un-controlled freeway driving. We then run two well known CAN bus anomaly detection techniques on the collect data. We find that neither technique flags messages corresponding to the RDA as anomalous, and we hypothesize that CAN bus detection techniques may not be suited for detecting this type of attack, thus motivating that the attack may be stealthy at the vehicular level as well.

\subsection{Data collection}
\label{subsec:data_collection}
First, we describe the data collection set up. \textit{CAN bus injection onto the ACC vehicle was approved by an institutional risk management entity, was in accordance with state law, and was collected only in an isolated area with trained vehicle safety operators. Benign data, which involved a non-expert vehicle operator, were collected in accordance with protocols approved by an Institutional Review Board (IRB).}

\begin{figure}
    \centering
    \includegraphics[width=0.95\columnwidth]{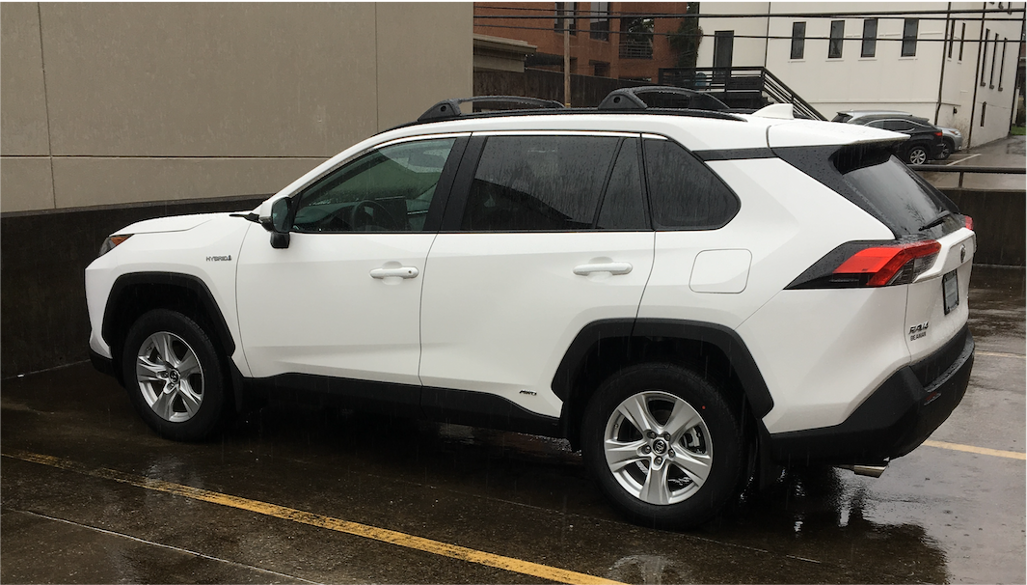}
    \caption{Toyota RAV4 Vehicle from which CAN bus data is collected.}
    \label{fig:vehicle}
\end{figure}

CAN bus data is collected from commercially-available Toyota Rav4 vehicles equipped with standard Toyota Safety Sense 2.0., as shown in Figure~\ref{fig:vehicle}. To access the CAN bus, we use a Black Panda manufactured by Comma.ai~\cite{santana2016learning}. The Black Panda is connected to a Raspberry Pi 4 via USB. Through open-source packages for the Robot Operating System integration \cite{canToRos} and CAN reading/writing \cite{libpanda}, the Pi is can both collect and send messages onto the vehicle's CAN bus on which ACC control is present.

Given the capacity to collect CAN bus messages from the ACC vehicle, we begin by observing its benign/normal behavior in live traffic. The vehicle is driven in active freeway traffic with ACC engaged, and so as not to bias the driving behavior at all, a non-author driver is employed who was not aware of the goal of the test and was simply asked to operate the vehicle using ACC, but take over temporarily if they felt it was unsafe to proceed. We collected 3 different continuous CAN bus message datasets from this scenario, each roughly 10 minutes long. We stress that, for this benign data set, CAN bus injections (\textit{viz.} RDAs) are not performed, and the vehicle is operated normally. All data was collected from the same roadway around the same time of day. 

Subsequently, we performed experiments in which CAN bus messages are injected on the CAN bus.
Given the nature of the proposed attack (RDAs) it would be unethical to perform this experiment in real traffic. As a result, we run these experiments on an isolated roadway. Additionally, due to safety concerns, experts operated the vehicle who were familiar with the experimental setup. For these experiments the driver initially began with the ACC operating at a set speed and subsequently issued CAN Bus injections of diverse magnitudes and durations, while periodically allowing the vehicle to return to normal behavior. Again, we collect 3 individual continuous CAN bus datasets each of roughly 10 minutes in length. It was our hope that these attack experiments be dramatic in their driving behavior, while the benign experiments are kept more modest, as this would give the CAN bus detection techniques the best chance to detect the attacks.

\subsection{Attack feasibility demonstration}

Given the data collected across the experiments in which messages are sent to the CAN bus from the compromised ECU (the Pi), we highlight a portion of one of the experiments that demonstrates the feasibility of the proposed attack.  

\begin{figure}
    \centering
    \includegraphics[width=0.95\columnwidth]{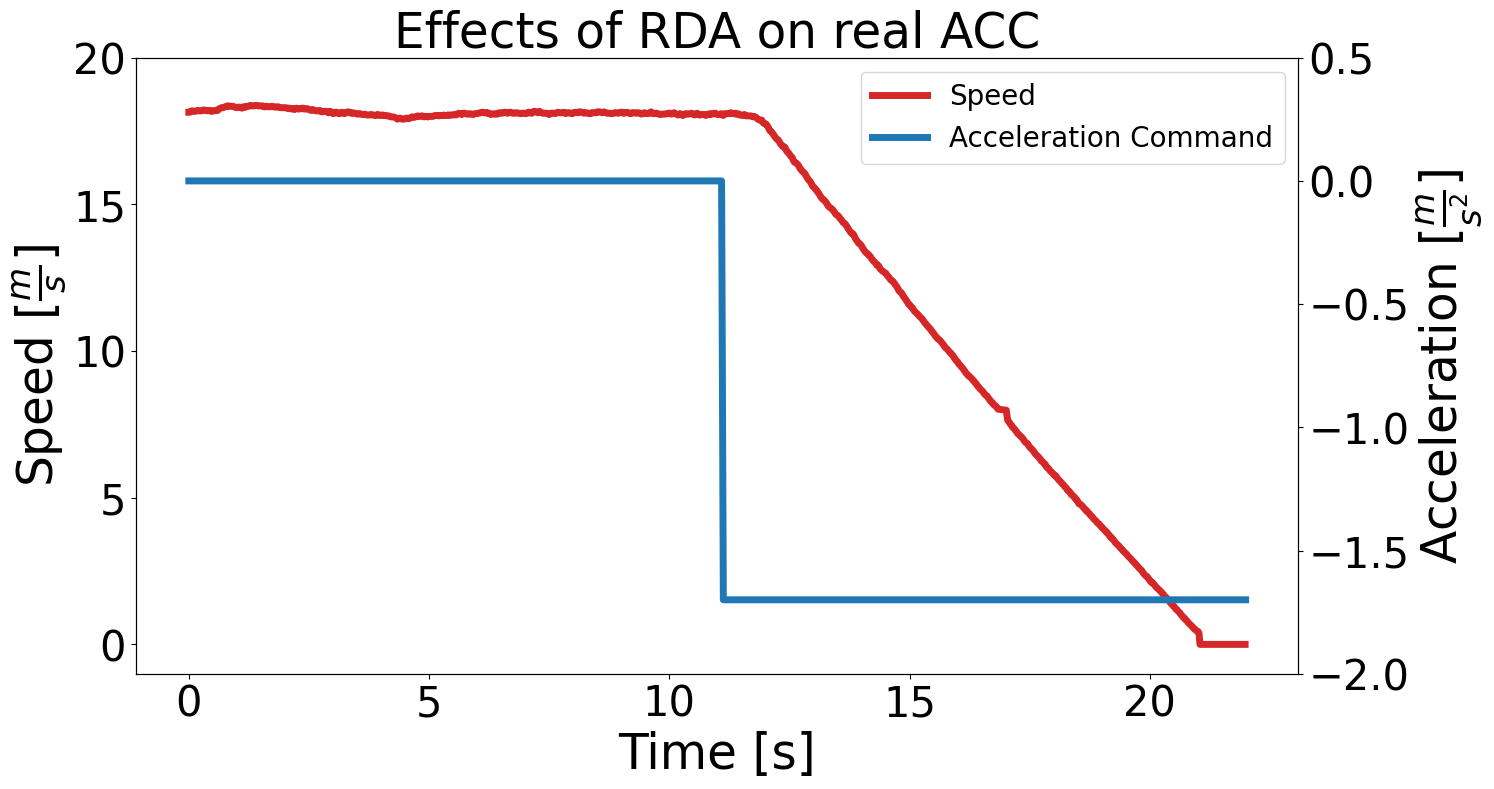}
    \caption{Effects of an attacker sending a constant deceleration command along a real vehicle's CAN bus via data overwriting.}
    \label{fig:decel_attack_real}
\end{figure}

First, the vehicle is brought to a steady operational speed of 40 mph using the ACC. No CAN bus messages are applied at this time. After 10 seconds of steady operation the system described in Section~\ref{subsec:data_collection} issues a steady command for the ACC to execute a deceleration of -1.7 $\frac{m}{s^{2}}$. The effect is that the vehicle steadily slows down at the prescribed deceleration rate, eventually reaching a complete stand still. The measured speed and the prescribed deceleration as measured on the vehicle's CAN bus are shown in Figure~\ref{fig:decel_attack_real}.

Throughout this time the ACC system was set to operate the system at 40 mph in the event that no vehicles are in front, which was the case throughout the entire experiment. Nonetheless, given commands to the ACC that instruct it to decelerate, despite the fact that this is not the expected operation of the system, the vehicle obeys the commands. This motivates the idea that if the ECU sending ACC acceleration commands were compromised with malicious software designed to execute RDAs it would indeed be able to execute the attack effectively on this vehicle. 

\subsection{Testing CAN bus anomaly detection techniques}\label{sec:test_CANBusDetect_real}
Having shown that an attacker can perform an RDA on a live vehicle if the ACC ECU is compromised in its software, we now investigate whether CAN bus anomaly detection techniques are able to identify this behavior as anomalous. To do this we employ CAN bus detection techniques on the collected datasets that learn from the benign data and test on the malicious.

Two state-of-the-art detection algorithms are tested. In~\cite{mirco2017anomaly} a classifier is trained by forming a transition matrix between two messages IDs if they are adjacent to one another in the observed training data. Subsequently, a pair of observed adjacent data are considered malicious if they had never previously been seen adjacent to one another. In~\cite{clinton2019automotive} a frequency based technique

The benign dataset is split into a training set and a validation set. In our experiments, the first $80\%$ of the benign data samples form the training set and the remaining $20\%$ are the validation set. Testing is done on all messages from the datasets containing injected messages.

\begin{table*}[]
    \centering
    \begin{tabular}{c|c c c c c c}
    \toprule
        Detection & Accuracy & True Positive & False Negative & False Positive & True Negative & AUC \\
        Method & & Rate & Rate & Rate & Rate &\\
        \midrule
        Transition Matrix & 0.95 & 0.0012 & 0.013 & 0.033 & 0.95 & 0.52 \\
        Frequency & 0.64 & 0.0 & 0.36 & 0.0 & 0.64 & 0.50 \\
        \bottomrule
    \end{tabular}
    \caption{Results for CAN bus anomaly detection on the proposed vehicle control attack.}
    \label{tab:CAN_detection}
\end{table*}

Metrics of success for these algorithms are shown in Table~\ref{tab:CAN_detection}. While the transition matrix based method has good accuracy, it fails to flag the anomalous messages, achieving effectively a 0\% true positive rate, and a 52\% AUC. The frequency based method is neither accurate, nor does it flag the malicious messages, achieving a 0\% true positive rate, and a 50\% AUC. In other words, both classifiers have a true positive rate that is so small that it can be said they completely fail to detect the RDA CAN bus injections. 

This result may seem surprising as CAN bus anomaly detection has received significant research attention recently. However, we propose that the failure to capture this attack is due to a difference in scope between our proposed attack and what CAN bus detection techniques aim to achieve. Primarily, CAN bus detection methods do not use the actual content of messages to perform detection. In other words, they are not model-based. Rather, most attempt to use characteristics of the timing on messages to identify when attacks are being executed, which allows these techniques to look across all possible CAN bus messages. In addition, since CAN bus detection techniques attempt to detect abnormal messages at the level of the entire communication network, it is computationally infeasible to run model-based techniques for all possible CAN bus messages.

While it is not possible to survey the entirety of CAN bus detection literature we suspect that even widely-cited detection techniques, such as~\cite{cho2016fingerprinting}, will not consider this attack in this work malicious. The most comparable attack outlined in~\cite{cho2016fingerprinting} to what we assume in our work is that of a strong fabrication attack, in which a compromised ECU is able to broadcast messages that are flagged as being from a different ECU, thus disrupting communication. In our attack, we assume rather that the data component of a CAN bus message being broadcast by the ECU is changed from the value it 'should' be, but that the same ECU is still sending the messages. The method in developed in \cite{cho2016fingerprinting} could not be tested in this work as it would require the real compromising of the ECU broadcasting ACC acceleration commands, rather than the read-delete-overwrite method that we are able to employ.  

Finally, it is worth noting that any CAN bus detection algorithm that are in fact run on the commercially available vehicle's which were tested in this work, do not succeed in flagging our experiments as malicious. While we do not have manufacturer level knowledge of whether or not detection strategies are truly being run on the tested vehicles, these experimental results suggest that current CAN bus detection methodologies are not suited for detecting this type of attack.

\section{Discussion}\label{sec:discussion}
\subsection{Improvements to traffic monitoring}
The lack of fidelity in the measurement of automotive traffic leaves it vulnerable to undetectable cyber-attacks as the pace of vehicle automation moves ahead of sensing technology. In particular, measurements techniques that only measure aggregate metrics on the traffic are not suited for finding individual malicious actors. GPS measurements pose a possible remedy, but are limited in the scope of their help because only a portion of traffic can be expected to have GPS devices active, meaning that it is very difficult, or even impossible, to contextualize a vehicle's movement with respect to whether it is responding to vehicles around it or engaged in a malicious manoeuvre. 

The two most promising paths forward for ensuring an ability to detect and act against cyber-attacks on AVs which target our transportation networks are: i) Ensuring that cyber-compromises of portions of the traffic flow are not possible, due to safe communication and software design on vehicles, and ii) advancing to observation of traffic flows which observe 100\% of vehicles along a network. While significant work has been and is being done on the first approach, supply-chain compromises to vehicular components represent a real and significant threat. In order to achieve the second approach to attack prevention either vehicles can be required to contain GPS units which are apart from the vehicular communication network, or monitoring infrastructure can be advanced to include overhead cameras along roadways. Both approaches would require significant support across both industries and governments, and could be very expensive. 

\subsection{Limitations and Future work}
Both attacks and defenses warrant further exploration. In terms of attack algorithms we explore only a naive (but still effective) attack that is unaware of the defense mechanism being employed. Further work could explore attacks that are designed to optimally affect traffic while staying stealthy, either in terms of the set of commands sent to the vehicle (in this case $a_\text{attack}$) or when an RDA should be executed (in this case this is done randomly). While we hope that attackers could not feasibly execute such optimized attacks, knowing their impact would help network operators assess the severity of the threat more fully.

A limitation of this work is that we do not assess the extent to which human operators in ACCs could act themselves as anomaly detectors. The extent to which this is true would require significant experimental work past the scope of this work, but may be worthwhile as future work. Importantly however, if the expectation for the operation of AVs is that eventually human operators/drivers are completely removed (i.e. completely autonomous vehicles), it would no longer be possible for the human to act as a supervisor, thus motivating the need for other detection methods.

It is possible that sufficiently well-crafted detection techniques could outperform those tested here to the point that they pose a legitimate means by which to fight attackers. This could consist of exploring more, and different, deep-learning architectures. Alternatively, it may be worthwhile to explore model-driven techniques, although such techniques will almost certainly require observational capacities which are significantly past current possibilities (i.e. complete observation). 

Finally, it would be of great interest to attempt to more completely measure the possible impact on a wider-spread execution of the attack. In this work we develop a high fidelity simulation environment for a real world highway environment, but only for a 3 kilometer stretch, upon which we quantify possible impacts. We provide a brief estimate of the cost incurred due to increased travel time, but this calculation could include other factors such as fuel consumption and safety. Extrapolating to the consequences of the attack at a national or international level would require significant modeling and data collection, but could provide vital information for agencies deciding what investments are feasible for defending against such attacks going into the future. Additionally, given the modeling/simulation based approach needed for this work, better and more detailed modeling approaches will lead to better understanding of the problem.

\section{Conclusions}\label{sec:conclusions}
In this work we explored the possibility that a cyber-compromise of the longitudinal control of ACC vehicles in a mixed traffic environment can degrade the performance of traffic overall. The concrete form of the attack is a random deceleration attack (RDA) in which the ACC is commanded to brake at a random time. The impact of this attack is quantified in a detailed/realistic traffic simulation environment, which is part of a broader open-source modeling software package (Anti-Flow) that we have developed for this topic. We find that this attack has different levels of impact in different traffic congestion regimes, with commuter metrics being significantly impacted in moderate congestion regimes, with average commuter speed being reduced by as much as 7\%. In high congestion regimes we find that throughput of the system, which is a system efficiency metric, can be degraded by up to 3\%, representing almost 300 fewer $\frac{veh}{hr}$ able to pass through the network.

We also develop a defense model, which involves anomaly detection on GPS measurements coming from mobile devices. We find that the attacks explored in simulation are stealthy to detection, meaning that the anomaly detection technique is unable to distinguish meaningfully between malicious/compromised vehicles and benign/normal vehicles. Despite this, the detection technique is able to correctly and reliably classify malicious vehicles if the attack is sufficiently strong. However, a counter-intuitive result is that in the context of especially strong attacks, many benign vehicles may also be labeled as malicious, due to the tendency for speed profiles of vehicles to be similar to one another. How readily strong attacks can be caught will depend on the extent to which a defender can handle high false positive rates.

Finally, we experimentally demonstrate on a live ACC vehicle that the proposed compromised at the vehicular level, which assumes an ability to overwrite acceleration commands, is able to cause the experimental ACC to execute an RDA. In particular, we are able to successfully command the vehicle to decelerate for over 10 seconds at a rate of $1.7 \frac{m}{s^2}$, representing an RDA as strong as those explored in simulation. This motivates that it is reasonable to think that if a supply-chain compromise of the type proposed were to happen, it is plausible that this attack could be executed. We also run data from the live experiments through two CAN bus detection techniques, both of which do not identify the overwritten acceleration commands as malicious, suggesting that the attack may even be stealthy at the vehicular level as well.

The general conclusion of this work is that current-day mixed autonomy traffic flows may be vulnerable to distributed cyber-compromise attacks which can significantly degrade transportation system performance while being stealthy to detection.

\bibliographystyle{plain}
\bibliography{refs}


\end{document}